\documentclass[fleqn,10pt]{wlscirep}
\usepackage[utf8]{inputenc}
\usepackage[T1]{fontenc}

\title{The Resurgence of the \textbf{$G(2)$} Group for the Strong Sector and the Emergence of Dark Matter}

\author[1,*]{Nicol\`o Masi}

\affil[1]{INFN \& Bologna University, Physics Department, Via Irnerio 46 Bologna, 40126, Italy}

\affil[*]{masin@bo.infn.it}

\begin{abstract}
$G(2)$ is the smallest exceptional group and it is the simplest and viable gauge group to minimally extend the strong interaction sector: $G(2)$ includes the group $SU(3)$ of Quantum Chromodynamics (QCD) as a maximal subgroup and it is equipped with six additional gluons that can acquire mass via a Higgs mechanism driven by a new Higgs particle and constitute dark matter. In this article I want to describe how the exceptional $G(2)$ group can be a physical gauge group, capable of extending the Standard Model (SM) of particles and including a versatile dark sector, which is compatible with experimental observations. In fact, due to its peculiar mathematical features, the group $G(2)$ manifests some complex features, not properly considered in literature, which guarantee its correct use in physics, as its $\{3\} \oplus \{\overline 3\}$ decompositions w.r.t. $SU(3)$ can acquire a complex structure. The resulting framework can be a solid Beyond Standard Model (BSM) solution for the dark matter (DM) problem, in the form of massive complex scalar glueballs, and it includes the proper color representations for quarks and leptons. Several quantum field theory features are discussed, like the $G(2)$ coupling constant running and its spectrum before and after the phase transition, along with all the DM astrophysical realizations, in order to present the unexpected potential of this gauge group.
\end{abstract}

\usepackage{latexsym}
\usepackage{amsmath}
\usepackage{amsfonts} 
\usepackage{bm}
\usepackage{textcomp}
\usepackage{graphicx}
\usepackage{adjustbox}
\usepackage{times}
\usepackage[square]{natbib}
\usepackage{deluxetable}
\usepackage[english]{babel}
\usepackage{wasysym}

\begin{document}
\flushbottom
\maketitle

\UseRawInputEncoding
\newcommand{\Z}{{\sf Z \!\!\! Z}}
\newcommand{\R}{{\sf I \!\! R}}
\newcommand{\C}{{\sf C \!\!\! C}}
\newcommand{\1}{{\sf 1 \!\! 1}}
\newcommand{\0}{{\sf 0 \!\! 0}}
\newcommand{\p}{\partial}
\newcommand{\Psibar}{\bar{\Psi}}
\newcommand{\kbra}[2]{|#1\rangle \langle #2|}
\renewcommand{\d}{\delta}
\renewcommand{\l}{\lambda}
\renewcommand{\L}{\Lambda}
\renewcommand{\b}{\beta}
\renewcommand{\a}{\alpha}
\newcommand{\lm}{\l_0}
\renewcommand{\k}{\kappa}
\renewcommand{\ni}{\noindent}
\newcommand{\g}{\gamma}
\newcommand{\n}{\nu}
\newcommand{\m}{\mu}
\renewcommand{\r}{\rho}
\renewcommand{\P}{\Phi}
\newcommand{\bx}{{\mathbf{x}}}
\newcommand{\by}{{\mathbf{y}}}
\newcommand{\bc}{{\mathbf{c}}}
\newcommand{\s}{\sigma}
\newcommand{\A}{{\cal A}}
\newcommand{\D}{{\Delta}}
\newcommand{\E}{{\cal E}}
\newcommand{\V}{{\cal V}}
\newcommand{\G}{{\large\cal G}}
\newcommand{\U}{{\large\cal U}}
\renewcommand{\S}{{\cal S}}
\newcommand{\vL}{\vec{L}}
\newcommand{\va}{\vec{\alpha}}
\newcommand{\vH}{\vec{H}}
\newcommand{\tS}{\tilde{S}}
\newcommand{\tU}{\widetilde{U}}
\newcommand{\J}{{\cal J}}
\renewcommand{\th}{\theta}
\renewcommand{\to}{\overline{\theta}}
\newcommand{\tm}{\theta^M}
\newcommand{\tp}{\theta^{ph}}
\newcommand{\tpo}{\overline{\theta}^{ph}}
\newcommand{\tpom}{\overline{\theta}^{ph}(\tm)}
\newcommand{\intpi}{\int^\pi_{-\pi}}
\newcommand{\e}{\epsilon}
\newcommand{\ep}{\varepsilon}
\newcommand{\vph}{\varphi}
\newcommand{\oh}{{\textstyle{\frac{1}{2}}}}
\newcommand{\oth}{{\textstyle{\frac{1}{3}}}}
\newcommand{\oq}{{\textstyle{\frac{1}{4}}}}
\newcommand{\ot}{{\textstyle{\frac{3}{2}}}}
\newcommand{\dg}{\dagger}
\newcommand{\non}{\nonumber}
\renewcommand{\t}{\tau}
\newcommand{\rf}[1]{(\ref{#1})}
\newcommand{\ra}{\rightarrow}
\newcommand{\pa}{\partial}
\newcommand{\ph}{\phi}
\newcommand{\phd}{\phi^\dagger}
\newcommand{\CR}{\nonumber \\}
\newcommand{\ww}{w}
\newcommand{\rra}{\right\rangle}
\newcommand{\lla}{\left\langle}

%%%%%%%%%%%%%%%%%%%%%%%%%%%%%%%%%%%%%%%%

\section{Introduction} \label{Introduction} \label{Intr}

As discussed in \citep{Masi2021}, a mathematical criterion to extend the Standard Model of particle physics from an algebraic conjecture was proposed: the symmetries of physical microscopic forces could originate from the automorphism groups of main Cayley--Dickson algebras, from complex numbers to octonions and sedenions. This correspondence led to a natural enlargement of the SM color sector, from a $SU(3)$ gauge group to an exceptional $G(2)$ group, broken by a Higgs mechanism: the broken symmetry produces an additional ensemble of massive $G(2)$--gluons, which is separated from the particle dynamics of the Standard Model.
$G(2)$ could represent the optimal gauge group to describe strong interaction and dark matter at the same time. Regardless of the mathematical process described in \citep{Masi2021} that led us to identify this new gauge group, I want to demonstrate here, as continuation of \citep{Masi2021}, that $G(2)$ is indeed a good group candidate for solving the dark matter problem, keeping the SM untouched. \\
$G(2)$ is always treated as a peculiar group not completely suitable for a physics theory: since all $G(2)$ representations discussed in physics literature are supposed to be real, the fundamental representation of the group is identical to its complex conjugate, so that $G(2)$ ``quarks'' and ``anti-quarks'' are conceptually indistinguishable. This induces obvious mathematical difficulties when dealing with SM particle fields and represents the technical obstacle for its use, except for applications in lattice QCD simulations (largely applied to simplify standard QCD computations \citep{Wellegehausen_2011,Maas:2012ts}, avoiding the so-called sign problem \citep{Wipf:2013vp} which afflicts $SU(3)$) and exotic $G(2)$ electroweak (EW) models \citep{Exceptionalg2EW}. \\
To expand the examination, a complex form of $G(2)$ was already identified by the great \'Elie Cartan, Friedrich Engel and Wilhelm Killing between the end of the 19th century and the beginning of the 20th century \citep{Cartan,Gantmacher1,Gantmacher2,AgricolaOldNew}: the analysis of $G(2)$ groups mathematical features leads to the comprehension that they can actually be exploited to build a SM gauge group, without affecting the known useful results and features of the real $G(2)$ theory. The emerging theory is a minimal extension of the SM with correct color charges, anticharges and singlet states, with massive complex vector fields which act as DM, and a promising behavior of the coupling running. The implications for the dark matter sector are rich and coherent with the DM phenomenology, offering a synthetic view of several BSM problems.\\
It must be stressed that this $G(2)$ proposal does not represent a Grand Unification Theory (GUT) for particle physics, because $G(2)$ does not include the EW sector: there is no direct coupling with $SU(2)$ and no induced proton decay processes by new EW physics. The following $G(2)$ theory broken by a Higgs mechanism can be understood as a counterpart of the EW symmetry breaking framework, producing few new heavy bosons, as the electroweak $W$ and $Z$ from the standard Higgs sector, having the role of dark matter. Thus this enlarged SM will have two complementary Higgs mechanisms for the EW and strong sectors: the usual one produces, from the overall EW symmetry, a massless photon and three massive interacting spin 1 bosons, the other the massless $SU(3)$ QCD gluons and the massive non--interacting spin 1 dark gluons. As for the visible world, this scenario is capable of producing aggregated states made up of $G(2)$ exceptional--colored particles \citep{Masi2021}. 
%%%%%%%%%%%%%%%%%%%%%%%%%%%%%%%%%%%%%%%%%%%%%%%%%%%%%%
%%%%%%%%%%%%%%%%%%%%%%%%%%%%%%%%%%%%%%%%%%%%%%%%%%%%%%
%%%%%%%%%%%%%%%%%%%%%%%%%%%%%%%%%%%%%%%%%%%%%%%%%%%%%%
%%%%%%%%%%%%%%%%%%%%%%%%%%%%%%%%%%%%%%%%%%%%%%%%%%%%%%
\section{The foundation of a $\textbf{G(2)}$ gauge theory for the strong sector of the Standard Model}\label{G2 QCD}
\subsection{Mathematical features}
\noindent In \citep{Masi2021} I showed that, in the real representation, $G(2)$ can be described as the automorphism group $\mathrm{Aut}(\mathbb{O})$ of the octonion algebra (also known as the Cayley numbers), which is a non-associative extension of the quaternions: the group $G(2)$ consists of all automorphisms of the octonions that fix the unit element. The group $G(2)$ is the simplest among the exceptional Lie groups \citep{Cacciatori:2009qu}; it is well known that the compact simple Lie groups are completely described by the following classes: $A_N(=SU(N+1)),  B_N(=SO(2N+1)),  C_N(=Sp(N)),  D_N(=SO(2N))$ and exceptional groups $G_2,  F_4,  E_6,  E_7,  E_8$, with $N=1,2,3,...$(for $D_N$, $N>2$) \citep{AdvModAlg}. Among them, $SU(2), SU(3), SO(4)$ and symplectic $Sp(1)$ have 3-dimensional irreducible representations and only $SU(3)$ between them has a complex triplet representation (this was one of the historical criteria to associate $SU(3)$ to the three color strong force, with quark states different from antiquarks states \citep{FondQCD}). There is only one non-Abelian simple compact Lie algebra of rank 1, \textit{i.e.} the one of $SO(3) \simeq SU(2) = Sp(1)$, which describes the weak force, whereas there are four of rank 2, which generate the groups $G(2)$, $SO(5) \simeq Sp(2)$, $SU(3)$ and $SO(4) \simeq SU(2) \otimes SU(2)$, with 14, 10, 8 and 6 generators, respectively \citep{Holland_2003}. \\
If we want to enlarge the QCD sector with a minimal rank 2 algebra to include dark matter, it is straightforward we have to choose $G(2)$ or $SO(5)$. The group $G(2)$, beside its intriguing relation with division algebras, is of particular interest because it has a trivial center, the identity, and it is its own universal covering group, meanwhile $SO(5)$ has $\mathbb{Z}_2$ as a center and it is not simply connected and its universal covering group is spin $Spin(5)$ \citep{Holland_2003}. None of $A_N,B_N,C_N,D_N$ groups is simultaneously simply connected and center free \citep{Hsiang_general}. The non-perturbative features are related to the topological objects that gauge field configurations can produce: for $G(2)$, the first and second homotopy groups are again trivial, while the third is the group of integers, hence it admits "instantons" \citep{thermoG22014}.
In addition, as all exceptional groups, $G(2)$ is anomaly free, that means safe from triangle anomalies, as shown by Georgi and Glashow \citep{GG_72,Gud_72}, and hence renormalizable. $G(2)$ subgroups are $SU(3)$, $SU(2)$ and $SU(2) \times SU(2)$. Exceptional groups always contain a $SU(3)$ subgroup and have $SU(3)$ singlets (leptons) and triplets (quarks) \citep{Stech1980}. For example, the exceptional group $E_6$ has been largely studied as a GUT \citep{Stech2004,Stech2008,Stech2012}, and $E_n$ in general are used in supergravity and string theory \citep{Book_Superstring,Super_Nath}. Proposing to enlarge QCD above the TeV scale and have the SM as a low energy theory is surely not an unprecedented nor odd idea: for example, modern composite Higgs theories \citep{Marzocca_2014,Da_Rold_2019,Cacciapaglia_2019} try to introduce (cosets) gauge groups beyond $SU(3)$, such as $SU(6)/SO(6)$, $SO(7)/SO(6)$ or $SO(5)/SO(4)$, dealing with multiple Higgs, strong composite states and dark matter candidates. \\
The key point to promote $G(2)$ to a viable gauge theory is that $G(2)$ representations w.r.t. the real compact $SU(3)$ group (with real adjoint representation of real dimension 8) are not necessarily real. It must be noted that, even though only the real compact representation has been study in literature, Cartan, Engel and Killing demonstrated there could be three versions of the $G(2)$ group \citep{Cartan,Gantmacher1,Gantmacher2,AgricolaOldNew}. In fact, $G(2)$ is basically the name of the group which refers to three simple Lie groups, a complex form, a compact real form and a split real form \citep{Bryant87,DRAPERFONTANALS_G2notes}: i) the complex Lie algebra $g(2)_{\mathbb{C}}$ (complex dimension 14, real dimension 28), with the compact form of $g(2)$ as the maximal subgroup, and representing its holomorphic complexification; ii) the Lie algebra of the compact form, which is 14-dimensional, the proper $\mathrm{Aut}(\mathbb{O})$, which is the one that has been mostly studied in literature and discussed in \citep{Masi2021}; iii) the Lie algebra of the non--compact split form, I will not discuss in the present dissertation. 
In standard calculations in physics, only the compact real form ii) has been used \citep{Pepe_2006,Masi2021}, with fundamental representation $\{7\}$ and adjoint $\{14\}$, decomposing w.r.t. $SU(3)$ into $\{7\} = \{3\} \oplus \{\overline 3\} \oplus \{1\}$ and $\{14\} = \{8\} \oplus \{3\} \oplus \{\overline 3\}$, respectively. The 14 generators are explicitly written in (1)--(5).
The "complexified" version of standard compact real $G(2)$ should have the same representations and the same decompositions w.r.t. the subgroup of the strong interaction to be safely used. We know \citep{Bryant87,AgricolaOldNew,DRAPERFONTANALS_G2notes} that $G(2)_{\mathbb{C}}$ is simple, connected and simply connected, and it is isomorphic to the group of algebra automorphisms of the "complex octonions"; its algebra is $g(2)_C = g(2) \times \mathbb{C}$. \\
One objection regarding the use of the real compact form of $G(2)$ is clearly stressed in \citep{Holland_2003}: if the $\{7\}$ fundamental fermion representation is real, it is identical to its complex conjugate (\textit{i.e.} self--conjugate and not \textit{reflexive}), so that $G(2)$ ``quarks'' and ``anti-quarks'' are conceptually indistinguishable. Then, in a pure unbroken $G(2)$ theory, we cannot deal with usual Dirac fermions but rather with Majorana particles: for real $G(2)$ representations, the $\{3\} \oplus \{\overline 3\}$ term of the fundamental representation could introduce a doubling of the color triplet and consequent issues in defining the EW transformation properties. So we have two ways to deal with this apparent issue: i) we can try to introduce a complex $G(2)$ representation which induces a complex (and so chiral) representation of $SU(3)$ or ii) we have to demonstrate that $\{3\} \oplus \{\overline 3\}$ linear representations can acquire a complex structure, although $\{7\}$ is real. Let's start from the first case. \\

\textit{i)} From a mathematical point of view, the congruence of the two compact and complex $G(2)$s is not so unexpected: for any semisimple Lie group generically called $G$, the finite--dimensional complex representations of a given complex form restrict to unitary (or more precisely, unitarizable) representations of its maximal compact subgroup $K$ \citep{Dong_complex, Hall}. In addition, there is an equivalence of symmetric monoidal categories between the category of finite--dimensional complex representations of $G$ and finite--dimensional unitary representations of its maximal compact subgroup $K$ \citep{BaezMonoidal}: this means that the tensor products of representations decompose in the same way. This general result can be applied taking $G$ to be the complex form $G(2)_{\mathbb{C}}$ and $K$ to be the compact real form that sits inside $G$ as a maximal compact subgroup, i.e. $G(2)$.
In details, if $K$ is a connected compact Lie group, then there exists a unique connected complex Lie group $G$ such that Lie algebras $g = k \times \mathbb{C}$ and $K$ is the maximal compact subgroup of $G$ itself. This is called the complexification of $K$ \citep{Dong_complex,Hall,Hsiang_general}. Treating a finite--dimensional holomorphic representation $\pi$ of $G$, this is holomorphic if and only if the associated Lie algebra representation of $g$ is complex linear, \textit{i.e.} $\pi(X + \textbf{i} Y) = \pi(X) + \textbf{i} \pi(Y)$, for $X, Y$ in $k$ algebra. If $G$ is simply connected, we can pass from the algebra $g$ to the $G$ group \citep{Hall}. In this case, every finite--dimensional \textit{rep} (I abbreviate for simplicity) of $K$ extends to a holomorphic representation of $G$ and every representation of $k$ algebra has a unique extension to a complex--linear representation of $g$, both denoted $\pi$ \citep{Hall}. Conversely, every holomorphic \textit{rep} of $G$ restricts to a representation of $K$. Furthermore, a \textit{rep} is irreducible for $G$ if and only if it is irreducible for $K$. So $G(2)$ and $G(2)_{\mathbb{C}}$ are equivalent.
In general, one has to also test the covering spaces of the group \citep{Hsiang_general,Hall}: \textit{e.g.} $G = SL(2,\mathbb{C})$ is the double cover of some different complex Lie groups $G'$ with the same Lie algebra, and similarly its maximal compact $K = SU(2)$ is the double cover of $K' = SO(3)$. This "covering" issue doesn't show up for $G(2)$s, which are their own universal covering groups: there is just one complex Lie group with $g_{\mathbb{C}}$ as its Lie algebra, and similarly for the compact form (see Appendix A for the thediscussion about not-necessarily-holomorphic representations).
This means, for the present exceptional group, that $G(2)_{\mathbb{C}}$ \textit{reps} can be generally treated in the same way we use the $G(2)$ ones. This resembles the development of the representations of the Lorentz group for the $G = SL(2,\mathbb{C})$ case, which is almost analogous \citep{Lorentz}.\\
So there is indeed a complete complementarity between the two complex and real $G(2)$, which grants the same treatment of complex and real irreducible (and adjoint) representations and their decompositions, even for not-necessarily-holomorphic irreducible \textit{reps}. Therefore the results for compact real $G(2)$ discussed in literature and especially in \citep{Masi2021} could be extended to the complexification $G(2)_{\mathbb{C}}$. This is relevant from a physical point of view, because it allows us to build up a correct gauge theory with the smallest complexified exceptional gauge group, but it leads to problematic features when one needs to break the group to recover the $SU(3)$ landscape pertaining to QCD. 
In fact, during the breaking to $SU(3)$, the complex form $G(2)_{\mathbb{C}}$ must be considered as a real Lie group, and it thus has real dimension 28 and not 14: besides the 8 $SU(3)$ generators (realizing the aforementioned real adjoint representation of $SU(3)$), $G(2)_{\mathbb{C}}$ possesses other 20 real generators w.r.t. its non--maximal subgroup $SU(3)$, which cannot be accounted by the 6 remaining 7x7 matrices alone, found in \citep{Masi2021} and reproposed in (\ref{due})--(\ref{cinque}). Indeed, the set of 14 complex matrices, as described a little later, provides a realization of the adjoint representation of the compact real form $G(2)$ (of real dimension 14), and not of the complex form $G(2)_{\mathbb{C}}$. So, even if it is promising and mathematically consistent, this $G(2)_{\mathbb{C}}$ solution should be discarded as non--minimal and source of further issues related to the corresponding Higgs sector. \\ 

%This is quite relevant from a physical point of view, because it allows us to build up a correct strong gauge theory with the smallest exceptional gauge group. From now on we denote by $G(2)_{\mathbb{C}}$ this physical complexified exceptional group which extends QCD $SU(3)$. \\
\textit{ii)} In the second case, we can try to keep the compact real $G(2)$, if and only if the $\{3\} \oplus \{\overline 3\}$ representations in the decompositions w.r.t. $SU(3)$ can acquire a complex structure, being not isomorphic to their dual. In fact, we should observe that, even if the fundamental representation $\{7\}$ of the 14-dimensional $G(2)$ is real, the resulting $\{3\}$ and $\{\overline 3\}$ representations of $SU(3)$ contained in the decompositions are complex and reciprocally conjugate: this is confirmed by the fact that the non symmetric coset $G(2)/SU(3)\cong S^6$\citep{Coset}, which on its tangent space locally reduces to the $\{3\}$ and $\{\overline 3\}$ linear representations of the isotropy group $SU(3)$ itself, has a complex structure $\textbf{J}$ \citep{Hall}, as demonstrated in \citep{DRAPERFONTANALS_G2notes}, proposition 5.2, or in \citep{AGRICOLA201875}, proposition 4.1, which allows to consider the 6--dimensional real vector space as a 3--dimensional complex space. Following a different approach, in \citep{Etesi_2015} this complex structure is interpreted as a classical ground state of a Yang--Mills--Higgs-like theory on $S^6$, such as a broken $G(2)$. These facts are obviously related to the problem of definition of whether the 6--sphere admits an actual complex structure \citep{Etesi_2015,Etesi_2015err,etesi2023complex,Tralle_2018,Konstantis_2018,Bryant2021}, which is not generally true, and it is linked to the properties of the corresponding Hodge number \citep{Angella_2018}. The realization of such an \textit{almost complex structure} granting the complex behavior of the $\{3\} \oplus \{\overline 3\}$ \textit{reps}, which is induced from octonions \citep{AGRICOLA201875}, is peculiar for the breaking of $G(2)$ into $SU(3)$ due to the $G(2)/SU(3)$ equivalence with $S^6$. In particular, restricting the 7--dimensional \textit{rep} of $G(2)$ to $SU(3)$ gives two dual--pair \textit{reps} of $SU(3)$ plus a trivial one. This is the solution we adopt for the following presentation.\\
We can exploit this important feature to develop a consistent physical description related to the strong sector. Clearly, this complex structure is recovered only w.r.t. $SU(3)$ and the unbroken $G(2)$ theory includes real $\{7\}$ \textit{reps} for particles, leading to fermions which have no reflexive (complex) \textit{reps} and cannot be distinguished from their antifermions, i.e. Majorana--like particles.
%%%%%%%%%%%%%%%%%%%%%%%%%%%%
As introduced in \citep{Masi2021}, to explicitly construct the complex matrices of real $G(2)$ in the fundamental representation, one can choose the first eight generators as \citep{Holland_2003,Pepe_2006}:
\begin{equation}
\label{su3gen}
\Lambda_a = \frac{1}{\sqrt{2}} \left( \begin{array}{ccc} \lambda_a & 0 & 0 \\ 
																													0 & \; -\lambda_a^* & 0\\ 0 & 0 &  0\
\end{array} \right).
\end{equation}
where $\lambda_a$ (with $a \in \{1,2,...,8\}$) are the Gell--Mann generators of $SU(3)$, which indeed is a subgroup of $G(2)$, with standard normalization relations $\mbox{Tr} \lambda_a \lambda_b = \mbox{Tr} \Lambda_a \Lambda_b = 2 \delta_{ab}$. $\Lambda_3$ and $\Lambda_8$ are diagonal and represent the Cartan generators w.r.t. $SU(3)$. \\
The remaining six generators can be found studying the roots and weight diagrams of the group. $G(2)$ has the only root system in which the angle $\pi/6$ appears between two roots. It is well known that the $SU(3)$ roots correspond to the long roots of $G(2)$. Starting from Patera's and Gunaydin's analysis \citep{Patera70,Gud73QuarkS,Beckers86,Gagnon88,Bincer93}, these additional generators can be written as complex matrices \citep{Zhong_weights,RevModPhys.34.1,Exceptionalg2EW,PhysRevD.99.116024,SM_higherdim_2003}:
\begin{equation}
\label{due}
\Lambda_{9}=\frac{1}{\sqrt{6}} 
\left(\begin{array}{ccc}0&-i\lambda_2&\sqrt{2}e_3 \\i\lambda_2&0&\sqrt{2}e_3 
\\ \sqrt{2}e_3^T & \sqrt{2}e_3^T&0\end{array}\right),
\Lambda_{10}=\frac{1}{\sqrt{6}} \left(\begin{array}{ccc}0&-\lambda_2&i\sqrt{2}e_3 
\\-\lambda_2&0&-i\sqrt{2}e_3 \\ -i\sqrt{2}e_3^T & i\sqrt{2}e_3^T&0\end{array}\right),
\end{equation}
\begin{equation}
\Lambda_{11}=\frac{1}{\sqrt{6}} 
\left(\begin{array}{ccc}0&i\lambda_5&\sqrt{2}e_2 \\-i\lambda_5&0&\sqrt{2}e_2 
\\ \sqrt{2}e_2^T & \sqrt{2}e_2^T&0\end{array}\right),
\Lambda_{12}=\frac{1}{\sqrt{6}} 
\left(\begin{array}{ccc}0&\lambda_5&i\sqrt{2}e_2 \\ \lambda_5&0&-i\sqrt{2}e_2 
\\ -i\sqrt{2}e_2^T & i\sqrt{2}e_2^T&0\end{array}\right),
\end{equation}
\begin{equation}
\Lambda_{13}=\frac{1}{\sqrt{6}} 
\left(\begin{array}{ccc}0&-i\lambda_7&\sqrt{2}e_1 \\i\lambda_7&0&\sqrt{2}e_1 
\\ \sqrt{2}e_1^T & \sqrt{2}e_1^T&0\end{array}\right),
\Lambda_{14}=\frac{1}{\sqrt{6}} \left(\begin{array}{ccc}0&-\lambda_7&i\sqrt{2}e_1 
\\-\lambda_7&0&-i\sqrt{2}e_1 
\\ -i\sqrt{2}e_1^T & i\sqrt{2}e_1^T&0\end{array}\right),
\end{equation}
\begin{equation}
\label{cinque}
e_1= \left(\begin{array}{c} 1 \\ 0 \\ 0\end{array}\right),
\,\,\,\,\,
e_2= \left(\begin{array}{c} 0 \\ 1 \\ 0\end{array}\right),
\,\,\,\,\,
e_3= \left(\begin{array}{c} 0 \\ 0 \\ 1\end{array}\right)\,\,\,.
\end{equation}
In the chosen basis of the generators it is manifest that, under $SU(3)$ subgroup transformations, as already anticipated, the 7-dimensional representation decomposes into \citep{Holland_2003,Pepe_2006,Masi2021}
\begin{equation}
\label{dec7}
\{7\} = \{3\} \oplus \{\overline 3\} \oplus \{1\} .
\end{equation}
This representation describes a $SU(3)$ fermionic color triplet $\{3\}$ (a quark--like particle), a $SU(3)$ anti-fermion $\{\overline 3\}$ and a $SU(3)$ singlet $\{1\}$. This color singlet can be interpreted as a lepton. So the previous fermionic representation that described fermions as a whole, without quark--lepton distinction, breaks into the usual picture of colored quark states and ``uncolored'' leptonic states. Under a $SU(2)\otimes U(1)$ group, the fundamental representation can be decomposed into doublets and singlets with their own hypercharge--like numbers, which resemble the fermionic left and right chiral states of the SM respectively:
\begin{equation}
\label{dec7ew}
\{7\} = (\{2\}_{1/2} \oplus \{1\}_{-1}) \oplus (\{\overline 2\}_{-1/2} \oplus \{1\}_{1}) \oplus \{1\}_{0} .
\end{equation}
Here it is worth to stress that the electroweak $SU(2)\otimes U(1)$ group of the SM is not embedded in this $G(2)$ strong group nor into its maximal subgroup $SU(3)$, due to the fact that $G(2)$ does not represent a unification theory for the overall $SU(3)\otimes SU(2)\otimes U(1)$ SM group, being $G(2)$ and $SU(2)\otimes U(1)$ independent interactions with independent Higgs mechanisms. The fact its EW decomposition does not exactly realize the SM fermions charges is the same as saying that there are no $SU(3)$ EW representations that contain the quark fields of the Standard Model, as discussed in the building of the exceptional electroweak sector in \citep{Exceptionalg2EW}. Furthermore it does not make sense to specify an overall fermion representation $G(2)\otimes SU(2)\otimes U(1)$ alternative to the SM one, because $G(2)$ does not necessarily coexist with the electroweak sector and indeed $G(2)$ hierarchically precedes it from an energy scale point of view, being broken at the EW scale, as it will be showed below.\\
The generators transform under the 14-dimensional adjoint representation of $G(2)$ \citep{Holland_2003,Pepe_2006}, which decomposes into \citep{Pepe_2007,Masi2021}
\begin{equation}
\label{dec14}
\{14\} = \{8\} \oplus \{3\} \oplus \{\overline 3\}.
\end{equation}
A $G(2)$ gauge theory, w.r.t. $SU(3)$, has colors, anticolors, color--singlets and 14 generators, so that it is characterized by 14 gluons, 8 of them transforming as ordinary gluons (as an octuplet of $SU(3)$), while the other six additional $G(2)$ gauge bosons separate into $\{3\}$ and $\{\overline 3 \}$, keeping the color quarks/antiquarks quantum numbers, being still vector bosons. 
%%%%%%%%%%%%%%%%%%%%%%%%%%%%%
\subsection{The $G(2)$-Higgs Yang--Mills theory}\label{YM}
A general Lagrangian for $G(2)$ Yang--Mills theory can be written as \citep{Masi2021}:
\begin{equation}
{\cal L}_{YM}[A] = -\frac{1}{2} \mbox{Tr} [G_{\mu\nu}^2],
\end{equation}
with the field strength $G_{\mu\nu} = \p_\mu A_\nu - \p_\nu A_\mu - i g_G [A_\mu,A_\nu]$ obtained from the vector potential $A_\mu(x) = A_\mu^a(x) \frac{\Lambda_a}{2}$, with $g_G$ a proper coupling constant for all the gauge bosons and $\Lambda_a$ the $G(2)$ generators. The associated canonical and "improved" energy--momentum tensors (EMT) for a pure Yang--Mills (YM) theory are \citep{EMT_2016}:
\begin{equation}
T_{\mu\nu}^{can}=2 \mbox{Tr} [-G_{\mu\rho}\p_\nu A^\rho + \frac{1}{4} \eta_{\mu\nu}G_{\rho\sigma}G^{\rho\sigma}],\qquad
T_{\mu\nu}^{imp}=2 \mbox{Tr} [G_{\mu\rho} G_{\nu}^\rho + \frac{1}{4} \eta_{\mu\nu}G_{\rho\sigma}G^{\rho\sigma}],
\end{equation}
where the improved one is not only conserved but also gauge invariant, symmetric and traceless.\\
$G(2)$ YM theory is asymptotically free like all non-Abelian $SU(N)$ gauge theories and, on the other hand, we expect confinement at low energies \citep{Pepe_2006}, with a different realization with respect to $SU(3)$, where gluons cannot screen quarks \citep{Masi2021}. The one--loop beta function for the running of the constant $g_G$, including gluons and fermions, can be found as usual as \citep{manoukian2016quantum,Czakon_2005,MACHACEK198383}:
\begin{equation}
\label{beta}
\beta(g_G)=\frac{g_{G}^3}{16 \pi^2}\beta_0=-\frac{g_{G}^3}{16 \pi^2}\left(\frac{11}{3}C_2(D) - s\frac{2}{3}n_{fl} S(R,f)\right)
\end{equation}
where $C_2(D)$ is the quadratic Casimir invariant of the adjoint representation $D$ of the group, $S(R,f)$ the Dynkin index of the fundamental irreducible $R$ representation for fermions, defined as a function of the generators $T_a(R)$ as $Tr[T_a(R) T_b(R)]=S(R,f)\delta_{ab}$, $n_{fl}$ the number of fermionic flavors and $s$ the spin factor, which is 2 for Dirac fermions and 1 for Majorana and Weyl fermions. For $SU(N)$, $C_2(D) = N$ and $S(R,f)$ are usually normalized to $1/2$, whereas for $G(2)$, defining $T_a(R) = \frac{\Lambda_a}{2}$, we can again obtain from the aforementioned normalization relation $S(R,f)_{G(2)} = 1/2$ and $C_2(D)_{G(2)}$ takes the value 4 \citep{Casimir2003} (which is naturally greater than $C_2(D) = 3$ for $SU(3)$ and equal to $SU(4)$). The number of flavors of $G(2)$ should be equal to the one of the SM, \textit{i.e.} six flavors of quarks.\\
Introducing a Higgs--like field in order to break $G(2)$ down to $SU(3)$, six of the 14 $G(2)$ ``gluons'' gain a mass proportional to the vacuum expectation value (vev) $w$ of the Higgs--like field, the other 8 $SU(3)$ gluons remaining untouched and massless. 
The Lagrangian of such a $G(2)$--Higgs model can be written as \citep{Masi2021}:
\begin{equation}
{\cal L}_{G_{2}H}[A,\Phi] = {\cal L}_{YM}[A] + (D_\mu \Phi)^2 - V(\Phi)
\end{equation}
where $\Phi(x) = (\Phi^1(x),\Phi^2(x),...,\Phi^7(x))$ is the Higgs--like field, $D_\mu \Phi = (\p_\mu - i g_G A_\mu) \Phi$ is the covariant derivative and $V(\Phi) = \lambda (\Phi^2 - w^2)^2$ the renormalizable quadratic scalar potential, with $\lambda > 0$. We can choose a simple vev like $\Phi_0=\frac{1}{\sqrt{2}}(0,0,0,0,0,0,w)$ to break $G(2)$ and re-obtain the familiar unbroken $SU(3)$ symmetry:
\begin{eqnarray}
\Lambda_{1-8} \Phi_0 = 0  & \textrm{(unbroken generators)}\\
\Lambda_{9-14} \Phi_0 \neq 0 & \textrm{(broken generators)}
\end{eqnarray}
This simple vev choice keeps the invariance w.r.t. the $G(2)$ long roots but not for the short roots. Plugging this scalar field vev into the square of the Higgs covariant derivative, we get the usual quadratic term in the gauge fields 
\begin{equation}\label{mmass}
g_G^2 \Phi_0^\dag \frac{\Lambda_a}{2} \frac{\Lambda_b}{2} \Phi_0 A_\mu^a(x) A^{\mu,b}(x) = \frac{1}{2} M_{ab} A_\mu^a(x) A^{\mu,b}(x)
\end{equation}
that gives the diagonal mass matrix $M_{ab}$ for the gauge bosons, of which we can use the aforementioned trace normalization relation, which Gell-Mann matrices and $G(2)$ generators share, to put the squared masses terms $g_G^2 w^2$ in evidence. 
This new scalar $\Phi$, which acquires a typical mass of $M_H = \sqrt{2\lambda} w$ from the expansion of the potential about its minimum \citep{ModernPP}, should be a different Higgs field w.r.t. the SM one, with a much higher vev, in order to disjoin massive gluons dynamics from SM one. Complementary, six massless Goldstone bosons are eaten and become the longitudinal components of the $G(2)$ vector gluons corresponding to the broken generators, which acquire the eigenvalue mass $M_G = g_G w$. The broken generators comprise the coset space 6--sphere $G(2)/SU(3)\cong S^6$.\\
The breaking of the $G(2)$ color string between two static $G(2)$ "quarks" happens due to the production of two triplets of $G(2)$ "gluons" which screen the quarks: $\{7\} \otimes \{14\}\otimes \{14\}\otimes \{14\} = \{1\} \oplus ...$represents a colorless $fGGG$ fermion-gluon state. When we switch on the interaction with the Higgs field, six $G(2)$ gluons acquire a mass thanks to the Higgs mechanism and the larger is $M_G$, the greater is the distance where string breaking occurs. When the expectation value of the Higgs--like field is sent to infinity, so that the massive $G(2)$ "gluons" are completely removed from the dynamics, also the string breaking scale is infinite. Thus the scenario of the usual $SU(3)$ string potential reappears. For small $w$ (on the order of $\Lambda_{QCD}$), on the other hand, the additional $G(2)$ ``gluons'' could be light and participate in the dynamics. Finally, for $w \rightarrow 0$ the Higgs mechanism disappears and we come back to $G(2)$. Only high $w$ values (with $w$ much greater that the SM Higgs vev) should be considered in order to realize a consistent dark matter scenario: in the $G(2)$ \textit{FIMP/SIMP} (Feebly Interacting Massive Particle/Strongly Interacting Massive Particle) cosmology for an exceptional scalar glueball described in \citep{Masi2021}, with the application of astrophysical observations constraints like the Bullet Cluster \citep{Robertson_2016}, the vev satisfies $w \geq 10^4$ GeV, which is completely consistent with a beyond SM phenomenology, with $w/M_H=O(1)$ and $M_S = M_{G\bar{G}}\simeq 2 g_G w$ for the mass of the scalar exceptional glueball made of two $G(2)$ massive gluons. 
According to the huge scale difference between the QCD scale and the supposed $G(2)$ one, the interactions might be suppressed in general. In fact, if we assume for example a small $\Lambda_{G(2)} \sim w \approx \text{20 TeV}$ value, it follows that $\Lambda_{QCD}/\Lambda_{G(2)}\approx \text{200 MeV}/\text{20 TeV}\approx 10^{-5}$. In addition, unlike GUT theories like the renowned Georgi--Glashow $SU(5)$ \citep{Burgess,ModernPP}, it must be emphasized again that $G(2)$ does not contain the EW $SU(2)\times U(1)$ sector and the new exceptional gluons do not easily interact with SM EW bosons nor fermions.\\
%%%%%%%%%%%%%%%%%%
\subsection{The coupling constant}\label{CC}
If one wants to study the behavior of the alpha constant associated to the $G(2)$ gauge theory which is broken into $SU(3)$, the beta function has to be completed with the scalar contribution, so that Eq.(\ref{beta}) $\beta_0$ becomes \citep{manoukian2016quantum,Czakon_2005,MACHACEK198383}
\begin{equation}
\beta_0=-\frac{11}{3}C_2(D) + s\frac{2}{3}n_{fl} S(R,f) + c\frac{1}{3}n_H S(R,S), 
\end{equation}
where $n_{fl}=6$, $n_H = 1$ is the number of the coupled Higgs--like scalars in the present theory, $c=1$ for a complex scalar and $c=1/2$ for a real scalar, and $S(R,S) = 1/2$ is the Dynkin index for scalars in the fundamental representation. 
To compare the coupling constant running with the others of the SM, computed w.r.t. the Z boson mass scale, the resulting $\alpha^{-1}_{G(2)}$ is:
\begin{equation}
\alpha^{-1}_{G(2)}(\mu) = const - \frac{\beta_{0,G(2)_{a,b}}}{2\pi} ln\left(\frac{\mu}{m_Z}\right)
\label{G2alpha}
\end{equation}
where $\mu$ is the energy scale in GeV and $\beta_{0,G(2)_a} = (-25/2, -151/12)$ for $s=1$ Weyl/Majorana fermions (with a complex and real scalar respectively), and $\beta_{0,G(2)_b} = (-21/2, -127/12)$ for $s=2$ Dirac fermions: this entirely defines the slope of the $G(2)$ alpha constant. The differences between the real and complex scalar cases are less than $1\%$, thus not significant for the behavior of the beta function. The constant term is not physical, because we cannot measure the $\alpha^{-1}_{G(2)}$ coupling at $m_Z$ energies being $G(2)$ already broken at the EW scale.
\begin{figure}[bht!]
	\centering
	\includegraphics[width=0.76\textwidth]{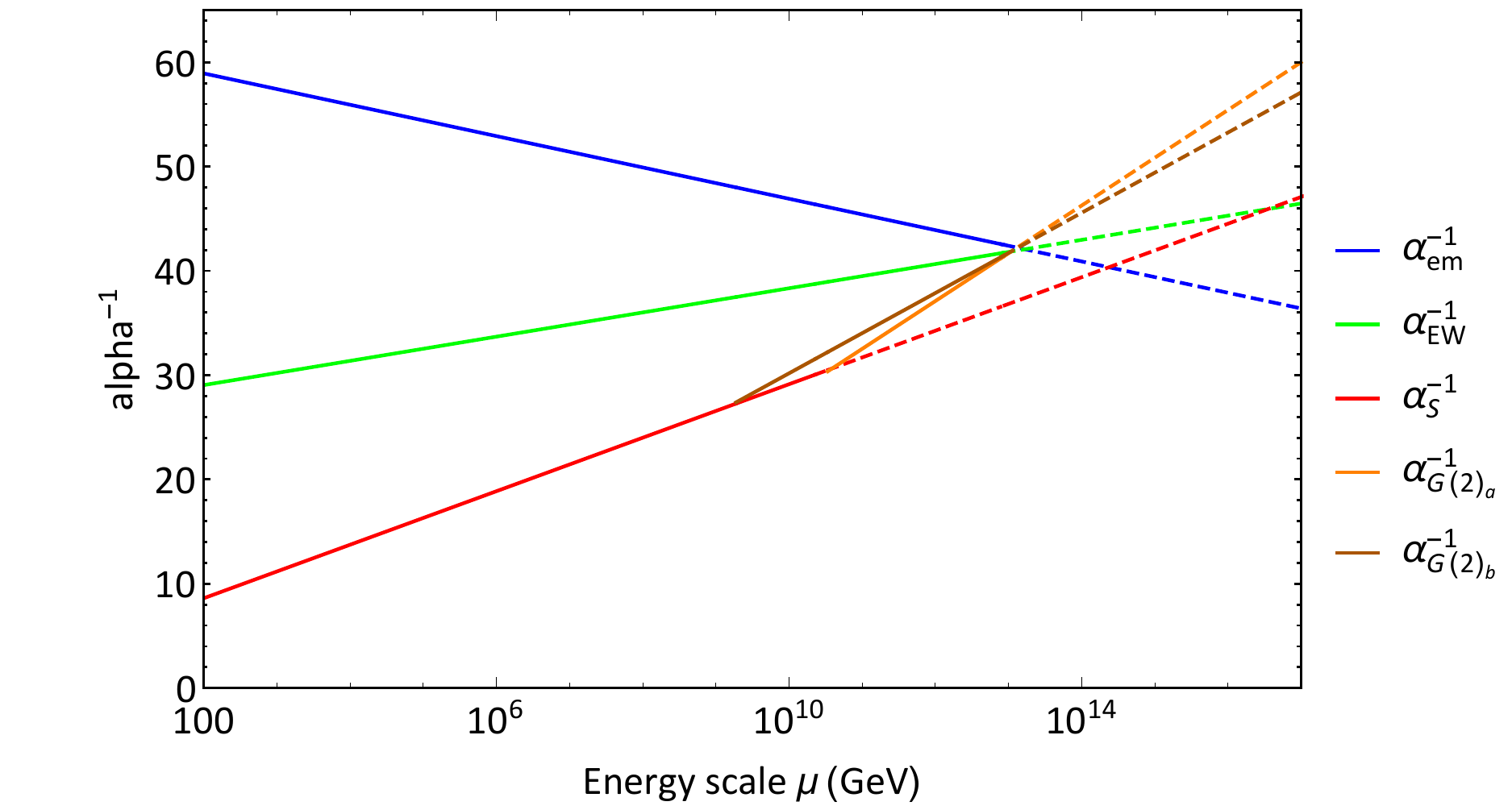}
	\caption{The running of the alpha constants for SM interactions and $G(2)$. The constant coefficient in Eq.(\ref{G2alpha}) is set to obtain the unification of $U(1)$, $SU(2)$ and $G(2)$ in the two aforementioned spin cases. The bins on the x-axis are logarithmic.}
	\label{run}
\end{figure}\\ 
%Both the spin alternatives are considered, because it is not obvious how to describe the $G(2)$ fermions in the $\{7\}$ representation, as discussed later.
The corresponding functions are displayed in Fig.1 (see the Appendix B for computational details), that shows the possibility to obtain the unification of the forces coupling constants via a transition between the $SU(3)$ regime and the $G(2)$ one: the slope changes due to the recovery of $G(2)$ symmetry, becoming steeper. This allows us to intercept the unification point of electromagnetism and weak force at about $10^{13}$ GeV, keeping the EW sector untouched. \\
One could obviously fix this $G(2)-SU(3)$ transition point at higher energies, passing the electroweak intersection point and invoking a more general GUT theory. But relying on this claim to unification, it is possible to determine a tentative transition in the strong sector which would occur approximately at $10^{10}$ GeV ($\alpha_{G(2)_a}\approx 1/30$) and $10^9$ GeV ($\alpha_{G(2)_b}\approx 1/27$), in the two spin cases. It is also interesting to notice that a complex scalar with a mass around the EeV scale, a quartic renormalizable potential and a small coupling with the SM Higgs has already been suggested in literature to stabilize the SM vacuum up to the Planck scale \citep{vevstability2012} (removing the metastability which occurs around $10^{10}$ GeV, due to the top quark large negative contribution to the Higgs quartic coupling \citep{vevstability2021,vevstabilityALL}). In \citep{vevstability2021} the SM vacuum stability is instead realized with a real scalar singlet. This metastability removal function could represent an important "added value" for the exotic Higgs which breaks $G(2)$ symmetry.\\
To explore the possibility of a gauge unification, even if the building of a GUT is beyond the present purpose, the smallest group which includes both $G(2)$ and a SM--like product $SU(2)\otimes U(1)$ would be the aforementioned famous exceptional $E_6$, which has $G(2)$ and $SU(3)\otimes G(2)$ as maximal special subalgebras, being $SU(2)\otimes U(1)$ the maximal regular subalgebra of $SU(3)$. It is worthy to note that unification approaches in literature using $E_6$ almost exclusively exploit its maximal regular subgroups $SU(3)\otimes SU(3)\otimes SU(3)$ \citep{Stech2004,Stech2012,PhysRevD.105.075021} and $SO(10)\otimes U(1)$ \citep{Schwichtenberg:2017xhv}, whereas its special subgroups have only recently been given attention \citep{Babu_2023}.\\
%%%%%%%%%%%%%%%%%
\subsection{The particle spectrum}
For what concerns the spectrum of a hypothetical $G(2)$ strong sector, the physics appears to be qualitatively similar to $SU(3)$ QCD \citep{masses2014,Masi2021}, but richer. 
In the spectrum of the unbroken $G(2)$ phase there are many more states beyond standard mesons and baryons: one-fermion-three-$G(2)$ gluons hybrid states (and, in general, the fermion confinement for one-fermion-$N$-$G(2)$ gluons, with $N\geq 3$), $(ff)$ difermions, $(ffff)$ tetrafermions and $(fffff)$ pentafermions, etc. If we take $f$ as a quark, states with baryon number 0 and 3 are somehow in common with QCD, whereas $B=1,2$, of $J=1/2$ hybrids and $J=0,1$ diquarks respectively, are $G(2)$ specific, as well as $B=4,5$ tetra--penta states. A tentative spectrum for the bosonic diquarks from real $G(2)$ lattice simulations has been proposed in \citep{masses2014}. $G(2)$ and $SU(3)$ also share glueballs states, for any numbers of $G(2)$ gluons (2 and 3 in the ground states) and hexafermions. The explicit decompositions of the products are reported in \citep{Masi2021}.\\
It should be noted that this "hadronic"--like categorization is peculiar because, before the breaking into $SU(3)$, these bound states do not distinguish between colored quark states and color--singlet lepton states nor between colored particle and antiparticles. Thus these  naturally neutral $G(2)$ states are made of generic Majorana fermions, not yet separated into quarks/antiquarks and leptons/antileptons, and these composite states are even more exotic than pure hadronic particles: they form a fermions plus glueballs ensemble with hybrid states. Again it must be stressed that they do not possess EW features yet and they should be singlet w.r.t. SM interactions. In such a sector, a sort of total absolute fermion number $F$ (as a baryon--lepton numbers $\left|B+L\right|$ sum) should be conserved, but not necessarily the separated baryon and lepton numbers after the symmetry breaking. For example, if one assumes $F=1$ in the $\{7\}$ $G(2)$ state, after the decomposition w.r.t. $SU(3)$ we have the same number of quarks and antiquarks charges ($\{3\}\oplus\{\overline 3\}$) with total baryon number $B=0$ and a lepton singlet with $L=\pm1$, with an overall conservation of the fermion number $F$ but not of $L$ alone. This could be a source of quantum numbers violation.\\
In this scenario, the new $G(2)$--breaking Higgs could be "dark" and entirely not capable of interacting with SM particles (so that the Yukawa couplings are zero), nor capable of giving the mass to fermions through a standard Yukawa coupling term of the type $Y \bar{L}\Phi R$ (with $Y$ the Yukawa coupling constant, $\bar{L}$ an antifermion left-handed $SU(2)$ doublet and $R$ a fermion right-handed $SU(2)$ singlet), which could be not viable because we still do not have quarks/antiquarks and leptons/antileptons fields separated. Even if the $G(2)$ Higgs is almost dark, as in some BSM theories \citep{Hiddensig,DarkHiggs17,King_2021,Stech_2014_Higgs}, residual interactions in the fundamental representation (\textit{i.e.} with $\{7\}$ Majorana fermions $\Psi=\Psi^{c}$), could allow a mass--like term of the type \citep{Srednicki,Zee} 
\begin{equation}
-\frac{1}{2}(m_M+y\Phi)(\Psi^{T}\textsl{C}\Psi+\bar{\Psi}\textsl{C}\bar{\Psi}^{T}) = -\frac{1}{2}\textsl{M}(\Psi^{T}\textsl{C}\Psi+\bar{\Psi}\textsl{C}\bar{\Psi}^{T}), 
\end{equation}
with $m_M$ a possible intrinsic Majorana mass, $y$ the coupling constant, $\textsl{M}$ the overall mass matrix, $\textsl{C}$ the charge conjugation matrix, $\bar{\Psi}=\Psi^{T}\textsl{C}$ and $\Psi=\textsl{C}\bar{\Psi}^{T}$ the Majorana conditions \citep{Srednicki} for the spinors: this could be an analogue of the SM Yukawa coupling for the real $G(2)$ fermions.\\
The symmetry breaking carried by the $G(2)$--Higgs gives the strong color $SU(3)$ feature to the $G(2)$ fermions lacking of separate baryon and lepton numbers. In this sense, the new Higgs does not act as a technicolor scalar leptoquark which transforms leptons into quarks and vice versa \citep{LQ1,LQ2,LQ3}, rather it induces the "universal separation" which splits the original generic fermion.
The usual SM Higgs symmetry breaking is a transition from EW massless eigenstates to massive measurable ones: it is described as an intrinsic mass--producing phase transition, which acts on flavors and chirality. On the other hand, the $G(2)$ breaking is the one which separates Majorana fermions into Dirac quarks/antiquarks and leptons, through the color feature of the fundamental $\{7\}=\{3\} \oplus \{\overline 3\} \oplus \{1\}$ representation. \\
When $G(2)$ transition occurs, as discussed above, fermion fields not necessarily have masses themselves (they are Majorana/Weyl fermions), because they cannot experience the SM Yukawa coupling which works at much lower energy scales, so all the singlet states constituents of the $G(2)$ spectrum could be massless, unless we introduce the aforementioned non--vanishing mass term for $G(2)$ Majorana fermions. 
Thus, if we could access the $G(2)$--Higgs phase transition (PT) scale, we could observe an almost massless phenomenology/dynamics, except for the scalar particle which triggers the phase transition, \textit{i.e.} the exotic Higgs, and the massive generators of the broken symmetry, \textit{i.e.} the dark exceptional gluons. But if we accept the existence of a \textit{mass gap} for a Yang--Mills theory \citep{Burgess} such as $G(2)$, there would always exist a non--zero mass state which represents the least massive particle predicted by the theory, \textit{e.g.} a glueball. In addition, regardless the mass term for the Majorana $G(2)$ fundamental particles, some composite fermionic states can acquire mass through the exchange of exceptional $\{14\}$ representation gluons within the bound system. The masses of the composite fermionic $G(2)$ particles described before might dynamically emerge from usual fermion and gluon kinetic energy, like in the proton \citep{pmass} case. The difference is that QCD--like fermionic condensates and anomalous contributions could not contribute because $G(2)$ fermions might be massless, there is no broken chiral symmetry and exceptional gauge theories are anomaly free \citep{Kou_2022}. Furthermore, being anomaly free, $G(2)$ is not affected by the $SU(3)$ strong CP problem and it does not introduce axion particles \citep{Peccei_2008} in the unbroken phase. \\
The aforementioned composite states could appear in extreme environments which recover the exact $G(2)$ symmetry, such as exotic neutron--like stars. In fact, collective manifestations of $G(2)$ exceptional matter could be distinguishable from QCD: for example, a $G(2)$--QCD neutron star could display a distinct behavior with respect to a $SU(3)$ neutron gas star, as shown in \citep{Hajizadeh_2017} for lattice purposes. $G(2)$ "real fields" matter could be a very interesting phase of matter to be studied for astrophysical purposes: once the $SU(3)$ QCD is no longer feasible and the involved temperature and pressure are sufficient to restore $G(2)$, the colored Dirac fermions cease to exist and they are replaced by their Majorana versions, which include color triplet, antitriplet and singlet. The resulting system could be treated as a realization of a "Majorana star" made of strongly interacting $G(2)$ particles.
In general, $SU(N)$ Yang--Mills theories with $N>3$ manifest first order deconfinement PT \citep{Lucini_2004,Nada:2015aia,Lucini_2014,Teper:2009uf,Panero_2009}, which are more markedly first order for increasing $N$. The peculiarities of the PT from lattice real $G(2)$ to $SU(3)$ have been extensively studied in \citep{Wellegehausen_2011,Nejad_2014,Cossu_2007,vonSmekal:2013qqa,thermoG22014}, confirming that $G(2)$ gauge theories have a finite--temperature PT mainly of first order and a similar but slightly discernable behavior with respect to $SU(N)$ \citep{thermoG22014}. In particular, in \citep{Wellegehausen:2011hY} the authors found the line of a first order PT connecting $G(2)$ and $SU(3)$ gluodynamics, whereas a line of second order PT between $G(2)$ deconfinement phase and the deconfined $SU(3)$.\\
%%%%%%%%%%%%%%%%%%%%%%%%%%%%%%%%
%\section{The exceptional dark sector and its manifestations: the boson star--black hole interplay}
\subsection{The exceptional glueballs: boson stars, black holes and Q--balls}
As discussed in \citep{Masi2021}, the $G(2)$ extension of the SM produces an \textit{exceptional} particle sector: if we move away the six $G(2)$ gluons from the dynamics, these bosons should be secluded and separated from the visible SM sector, without experimentally accessible EW interactions and extreme energies should be mandatory to access the $G(2)$ string breaking. 
In addition, $G(2)$ gluons are electrically neutral and immune to interactions with photons and weak $W$, $Z$ bosons at tree level. Another advantage of a $G(2)$ broken theory is that no additional families are added to the Standard Model, unlike $SU(N)$ theories: this seems a good scenario for a cold dark matter (CDM) theory (cold means non-relativistic and refers to the standard Lambda-CDM cosmological model), once provided stable or long-lived candidate. 
The \textit{dark} massive gluons from the $G(2)$ symmetry breaking can form dark glueballs constituted by two (or multiples) $G(2)$ gluons, with integer total angular momentum $J = 0, 2$ for the simplest 2--gluons balls, in which the massive constituents posses the color/anticolor features. To ensure the stability of the lightest $J^{PC}=0^{++}$ "pion--like" state at least, in \citep{Masi2021} few alternatives, besides the trivial condition $M_H>M_{GG}$ to avoid the decay into the exotic Higgs, have been proposed: i) an accidental symmetry, \textit{i.e.} a conserved additive gluon number $\Gamma$ to prevent the decay into mesons (like the baryon number $B$ for protons); ii) a global discrete $Z_2$ or $Z_n$ symmetry (or a continuous $U(1)$ symmetry), with DM which is odd under the new symmetry while SM fields are assumed to be even; iii) a $G$--parity conservation for a generic Yang--Mills theory, to prevent decay into $G$--even SM particles, unlike pions.\\
For such a dark sector a \textit{Dark freeze-out} cosmological scenario has been proposed in \citep{Masi2021}, for which DM reaches an equilibrium heat bath within the dark sector itself, never interacting with SM particles. The two--gluons $G(2)$ scalar glueball $S$ is \textit{ab initio} produced out of thermal equilibrium, for instance by a heavy mediator decay, such as the new $G(2)$--Higgs ($H\rightarrow DM DM$): this is viable if the coupling between DM and the heavy Higgs is sufficiently weak, realizing an \textit{exotic--Higgs portal DM} \citep{Masi2021,Bernal_2019}, which is described by: 
\begin{equation}\label{cosmology}
V(\Phi,S) = V(\Phi) + \frac{m^2}{2}S^2+\frac{\lambda_4}{4} S^4+\frac{\lambda_{HS}}{2} \Phi^2 S^2
\end{equation}
where $\lambda_4$ is the quartic self-interaction strength, $\lambda_{HS}$ is the heavy Higgs--scalar glueball $S$ coupling and $V(\Phi)$ is the exotic Higgs potential described in section \ref{YM}; ${M_S}^2=m^2+\lambda_{HS} w^2 /2$ can be defined as the total mass after the $G(2)$ symmetry breaking. The mass scale $m$ is already set by the new Higgs \textit{vev}, as $M_G = g_G w$ and $m \simeq 2 g_G w$. The dark matter mass is then defined as:
\begin{equation}
M_S=2 w\sqrt{g^2_{G}+\lambda_{HS}/8}.
\end{equation}
The energy scale of the DM mass is set by the intensity of the fundamental couplings $g^2_{G}=4\pi\alpha_{G(2)}$ and $\lambda_{HS}$. If $\lambda_{HS}$ is dominant but sufficiently small ($\lambda_{HS} \leq 10^{-7}$ \citep{Bernal_2017}), a \textit{FIMP} cosmology via a \textit{freeze-in} mechanism \citep{Hall_2010} is achieved, the DM mass is of the order of $w \sqrt{\lambda_{HS}} $ and one can recover a correct dark matter relic abundance $\Omega h^2 \simeq 0.12$ \cite{Aghanim:2018eyx}. In addition, it is possible to estimate the lower bound $w \geq 10^4$ GeV \citep{Masi2021}, obtained via the DM--DM scattering cross section limit from the Bullet Cluster \citep{Robertson_2016,Eby:2015hsq,Bernal_2019}, as already mentioned: for high self-interaction $\lambda_4 \sim 10$, $\lambda_{HS}\sim 10^{-(10\div 9)}$ and $w/M_H=O(1)$, the resulting scalar glueball mass is $M_S\sim 10^{-(5\div 4)} M_H$, naturally leading to a DM mass in the GeV region \citep{Bernal_2016}. Otherwise, if $g_G$ dominates and it is defined by the unification attempt of the alpha constants in section \ref{CC}, $g_G\approx 0.6\div 0.7$ sets the DM mass to the $w$ scale, which is roughly $\geq 10^9$ GeV. No unitary upper bounds for non--thermal DM can be universally achieved, unlike the $O(100)$ TeV limit for the usual WIMP thermal case from \textit{s}--wave annihilation \citep{Griest:1989wd}. \\
As anticipated, the \textit{FIMP} scenario can change if $\lambda_4$ is large enough ($\lambda_4 > 10^{-3}$ \citep{Bernal_2016}), \textit{i.e.} if scalar self-interactions are active: the DM particles, initially produced by the previous mechanism, may thermalize among themselves, due to number-changing processes (generic $n_{DM}\rightarrow n'_{DM}$ processes), which reduce the average temperature of DM particles and increase the number density until equilibrium is reached. The resulting relic abundance could therefore change even though the coupling between the visible and dark sector is absent. This could be the case of a \textit{SIMP} scenario with a dark freeze-out mechanism \citep{Bernal_2017,Bernal_2016, Bernal_2019}. See \citep{Masi2021} for details and insights about the realization of such a \textit{FIMP/SIMP} cosmology.\\ 
For negligible thermal production of DM particles, DM should come from a non-thermal mechanism, leading to a Super-WIMP (SWIMP)-like scenario, for example through a direct DM--producing primordial field decay, like the inflaton \citep{Allahverdi:2002nb,Almeida:2018oid,delaMacorra:2012sb}, if the heavy Higgs scalar responsible for the $G(2)\rightarrow SU(3)$ transition is identified as the inflaton field itself. In fact, heavy DM can be entirely produced during the reheating era \citep{Garcia_2020,Haque_2023}, which is the phase that bridges inflation and the standard Big Bang, repopulating the Universe. In this \textit{inflaton--portal} case \citep{Heurtier:2019eou}, the glueball abundance is basically fixed by few parameters, \textit{i.e.} the reheating temperature $T_{rh}$, the DM mass, the inflaton mass and its branching ratio $B_S$ into $G(2)$ gluons/scalar glueball, and it can be approximated to \citep{Allahverdi:2002nb}:\\
\begin{equation}
\Omega_{S}h^2 \simeq 2 \times 10^{8} B_S (M_S/M_H) \frac{T_{rh}}{\rm GeV}
\end{equation}
For a cosmological reheating above the GeV scale, the $B_S (M_S/M_H)$ factor should be quite small, \textit{i.e.} $< 10^{-9}$. 
As corroborated by recent and sophisticated calculations, the inflaton--portal scenario works well especially for large DM masses, between the weak scale and the PeV scale \citep{Almeida:2018oid}, and for extremely decoupled EeV$\div 10$ EeV candidates \citep{Heurtier:2019eou} and inflaton mass of $O(10^{13})$ GeV (large--field inflation models), which could be a case similar to the present heavy $G(2)$ glueballs one. Since the inflaton mass could be rather large as compared to the visible sector, interactions between DM and SM particles should be extremely suppressed, preventing thermal equilibrium, the two sectors being highly decoupled once they get populated by inflaton decay. The reheating will give rise to different temperatures for the two baths, according to the decay branching ratios of inflaton \citep{Heurtier:2019eou}. In general, the total DM relic abundance can be the sum of a freeze-in production and the dominant inflaton decay \citep{Garcia_2020}, also including gravity-mediated decay processes \citep{Haque_2023}, eventually modified by a \textit{SIMP} dark freeze-out \citep{Heikinheimo_2017,Choi_2017}.\\

Hereafter the boson star--black hole interplay is discussed. For what concerns the astrophysical manifestations of the $G(2)$ dark matter, exceptional glueballs could form boson gases or could collapse or aggregate into extended objects, which could populate the so--called dark halo. Our glueball scalar field $S$ evolving in the General Relativity framework can be described by the Einstein-Klein-Gordon (EKG) action,
\begin{equation}
	\label{EKG}
	\mathcal{S} = \int \mathrm{d}^4x\sqrt{-g} \left[\frac{R}{16\pi G} - g^{\mu\nu}\partial^\mu\bar S\partial_\nu S - V(S)\right]\,,
\end{equation}
where $V(S)$ is the bosonic potential, $\bar S$ the complex conjugate for a complex scalar, $R$ is the Ricci scalar, $g^{\mu\nu}$ is the metric of the space--time, $g$ its determinant and $G$ is Newton's gravitational costant. The variation of the action with respect to the metric leads to the related Einstein equations \citep{Liebling_2017,Cardoso_2019} for static, spherically--symmetric geometries. The key ingredients to build up stellar objects with non--light bosons are mainly the magnitude of the quartic order self-interaction $\lambda_4$ of the constitutive boson and its mass, given a generic quartic potential $V({\left|S\right|}^2)=\frac{M^2_{S}}{2}{\left|S\right|}^2+\frac{\lambda_4}{4}{\left|S\right|}^4$ for the complex scalar $S$, like the one in (\ref{cosmology}): repulsive self-interactions ($\lambda_4>0$) can balance gravity and give rise to very dense boson stars (BS) configurations with structured layers \citep{Schunck_2003}: the stress--energy tensor of a BS is in general anisotropic and the macroscopic scalar radial pressure works against gravity. Also rotating BS solutions do exist \citep{Adam_2022}, and they are expected to be possibly stable in the large coupling regime $\lambda_4 \gg M^2_{S}$ of the quartic self--interaction \citep{Vaglio_2022}. The maximum mass for such an object, the corresponding boson star radius and the critical particle number \citep{Chavanis_2012,Hertzberg_2021,Liebling_2017,Schunck_2003,Masi2021} are
\begin{equation}
M^{max}_{BS}\sim {\sqrt{\lambda_4}\,M_{Pl}^3\over M^2_{S}/{\rm GeV}^2}, \qquad
R^{max}_{BS}\sim {\sqrt{\lambda_4}\,\times 10\, {\rm km} \over M^2_{S}/{\rm GeV}^2}, \qquad
N^{max}_{BS}\sim {\sqrt{\lambda_4}\,M_{Pl}^3\over M^3_{S}/{\rm GeV}^3}
\label{Maxmass}
\end{equation}
where $M_{Pl}\equiv1/\sqrt{G}\approx 1.2\times 10^{19} GeV \sim M_{\odot}^{1/3}$ is the Planck mass and $M_{\odot}$ the solar mass, and $G$ is Newton's gravitational constant. In \citep{Masi2021}, an approximation to $SU(N = 4)$ of the quartic coupling of the type \citep{Hertzberg_2021} $\lambda_4 \sim (4\pi)^2/N^2 \approx \pi^2$ was proposed for $G(2)$. Such a boson star can potentially mimic a stellar black hole (BH) or a neutron star as a gravitational waves source but it could be in principle discriminated studying tidal perturbations of inspiraling binary systems \citep{BSdiscrim2017}, tidal effects of the gas clouds around the boson star \citep{BSdiscrim2021}, extrapolating the spin of the merger remnant \citep{BSdiscrim2021b}, or from peculiar universal relations for rotating BS between moment of inertia, angular momentum and quadrupole moment \citep{Adam_2022}. Interestingly, in \citep{SONI2017379} the authors found that the radius to mass ratio for a $SU(N)$ scalar glueball dark matter star is always larger than the Schwarzschild radius of a black hole with equal mass. Furthermore, a BS can be transparent, i.e. visible matter can accumulate into the center of the star or pass through the BS interior \citep{Schunck_2003}. So it might be possible to detect a BS also from its radiation emission and the associated gravitational redshift and gravitational lensing effects. On a larger scale, BS can even mimic a supermassive black hole (SMBH) in an active galactic nucleus \citep{Schunck_2003}.
All these observations strongly depend on the choice of the scalar potential, which could be changed to the giant boson halo, Log, Sine-Gordon and Cosh-Gordon or Liouville BS models (see \citep{Schunck_2003} for a complete review of BS physics): here the discussion is restricted to the standard Colpi--Shapiro--Wasserman quartic potential, which is sufficient and coherent with the minimal Higgs--portal case covered so far. Obviously, the feasibility of a star made of self-interacting scalars also depends on a correct estimate of the possible $3\rightarrow1$, $3\rightarrow2$ and $4\rightarrow2$ number changing annihilation processes inside the star \citep{Hertzberg_2021}, which in turn depend on the symmetries (\textit{e.g.} $Z_n$ or a global $U(1)$ conserving the particle number) of the exotic sector.\\
From the previous equation (\ref{Maxmass}), for a $M_S\sim 10$ TeV scalar, the resulting object carries a small mass $M_{max}\lesssim 10^{-8}\sqrt{\lambda_4} M_{\odot}$ w.r.t. usual stars, about one hundredth of the Earth mass and a sub--millimeter radius for $\lambda_4 \sim 1 \div 10$ (where the perturbativity limit $\lambda_4<4\pi$ must be fulfilled anyway \citep{Masi2021}), making it a "marble" of dark glueballs. Instead, for a 1 GeV scalar, $M^{max}_{BS}\lesssim \sqrt{\lambda_4} M_{\odot}$ and $R^{max}_{BS}\lesssim \sqrt{\lambda_4}\,\times 10\, {\rm km}$ stand, producing a solar mass object with a neutron star--like radius. Both the aforementioned $M_{max}$ ranges, \textit{i.e.} $10^{-8}M_{\odot}$ and few $M_{\odot}$, fall into mass windows with weak constraints for MACHOs from up-to-date microlensing analysis \citep{Schunck_2003, Calcino:2018mwh,Brandt:2016aco}. As stressed in \citep{Hertzberg_2021}, at the maximum mass the radius is slightly larger than the corresponding Schwarzschild radius and, for $\lambda_4 = O(1)$, the relation is similar to the Chandrasekhar mass for white dwarf stars $M_{wd}\approx M_{Pl}^3/m^2_{p}$; for the so--called mini boson stars, with no scalar interaction potential, the maximum mass is quite smaller and scales with $\sim M_{Pl}^2/M_{S}$.\\ 
So, the highest is the \textit{vev} of the $G(2)$--Higgs, the smallest is the resulting scalar glueball star. For instance, if we consider the scenario in Fig.1 and we take $w$ as an intersection point between $SU(3)$ and $G(2)$, we can obtain an estimate for the exceptional gluons mass and compute the effective radius of a boson star made of exceptional complex scalar glueballs: choosing the lowest energy case $\Lambda_{G(2)_b} \sim w\approx 10^9$ GeV, with $\alpha_{G(2)_b}\approx 1/27$ (the same reasoning can be done for $\Lambda_{G(2)_a}$), the consequent scalar mass is 
\begin{equation}
M_S\approx 2 g_G w = 2 w \sqrt{4\pi\alpha_{G(2)_b}} = 1.4 w \approx 10^9 {\rm GeV}.
\end{equation} 
For such an extreme particle mass, the star maximum radius is comparable to the one of a nucleus. To check if this incredibly tiny object might approach a black hole radius, we can qualitatively compute its Schwarzschild radius starting from a single glueball: $R_S [M\sim 10^9 {\rm GeV}] = 2 M G/c^2 = 2.6 \times 10^{-45}{\rm m}$. Therefore, if the star collects more than approximately $10^{31}$ exceptional glueballs (a number which is extremely lower than the number of protons in the Sun, for comparison), a nucleus--size event horizon $R_S \approx R^{max}_{BS}$ can be reached and the $G(2)$ glueball--star starts being a BH. Such an exceptional star, with mass of the order of $10^{-18} M_{\odot}$ and EeV constituents, becoming a BH, could play the role of a light primordial black hole (PBH) \citep{PBHDomingo2021}: in fact, the condition $M_{PBH} \gtrsim 10^{-(19\div 18)} M_{\odot}$ ensures the PBH will not evaporate in a time--scale comparable with the age of the Universe \citep{PBHmass2023}. In general, regardless the features of the glueball star, it is a priori always possible to create a population of dark matter--based primordial black holes if the very early Universe is born with a large--enough--magnitude density fluctuation so that it itself rapidly collapses \citep{PhysRevD.88.084051}, leading to the formation of primordial black holes well prior to the formation of any stars. These microscopic BHs originating from exceptional dark matter could merge together during the evolution of the Universe itself, producing larger and larger objects, being the seeds for more massive BHs, or even supermassive black holes \citep{BHseeds2012,BHseeds2021}, if the primordial merging rate is sufficiently high: even if part of the DM abundance is subtracted to form SMBH, it is true that the mass of a SMBH at the center of a spiral galaxy is usually only one $\permil$ of the mass of the hosting DM halo \citep{SMBHhalo1,SMBHhalo2}. This DM--BH equivalence and interplay scenario is easily achievable for very high $w$ and consequent DM mass. It must be stressed in this special case \emph{it is dark matter that produces PBHs rather than PBHs making up DM} \citep{PBHasDM2016}. So DM can be initially manifested as exceptional glueball particles, small boson stars or light PBHs in a complicated mixture. According to the fact that baryons can affect the DM halo through their own gravity and by modifying the gravitational potential of the galaxy, these tiny objects can be subsequently swept away from the galactic center via stellar feedback and DM heating \citep{DMheat1,DMheat2,DMheat3,DMheat5}, producing the "cored halo" we observe today \citep{DMheat4,DMheat6}.\\
Luckily, for a few years we can take advantage of both electromagnetic and gravitational waves astronomy as powerful probes to discriminate compact objects as a function of their "compactness" or "closeness" to a black hole (a quantity related to the mass to radius ratio $M/R$), "shadows" and gravitational waves emission \citep{Cardoso_2019}. For example, a hypothetical binary $SU(N)$ gluons star could be disentangled from a binary black hole system, due to possible differences in the gravitational wave frequency and amplitude, as demonstrated in \citep{SONI2017379}. As stressed before, it is possible that exceptional $G(2)$ DM should have different types of manifestations, \textit{i.e.} boson gases, boson stars and BHs, at the same time, as a function of its evolution history.
\\

%%%%%%%%%%%%%%%%%%%%%%%%
Another fundamental possibility for astrophysical macrostates built up via $G(2)$ glueballs not discussed in \citep{Masi2021}, is the \textit{Q--ball} solution, a type of non-topological soliton. This is a localized field configuration with a stability guaranteed by a conserved "charge", \textit{i.e.} the particle number \citep{shnir_2018}: the soliton has lower energy per unit charge than any other configuration. The soliton is approximately spherically symmetric. To build up a Q--ball, an attractive quartic self--interaction potential is needed, opposite to the boson star case, so that $\lambda_4<0$, along with a sextic repulsive potential term \citep{WickCutkosky_2017,UnderstandingThinWall_2021,Polynom_Pot_2022,Soliton_2022,Exited_Qb_2022}, under a $U(1)$ symmetry: the resulting Q--ball is a finite--sized object containing a large number of complex--valued scalar bosons, which is stable against fission into smaller objects and against "evaporation" via emission of individual particles, because, due to the attractive interaction, it is the lowest-energy configuration of that number of particles \citep{shnir_2018}. Q--balls can mimic solar masses objects like boson stars and BHs (or even SMBHs \citep{SupeMBH_2014,SMBH_Gb_2016}), and they have rich cosmological implications \citep{Bai_2022}. Eventually, the Q--ball can gravitationally capture baryonic matter, granting a ultrahigh density and pressure environment for the formation of radio sources or even black holes \citep{SMBH_Gb_2016}. In a very recent work \citep{PhysRevD.109.023003} their gravitational lensing features and mass--radius relations have been studied in details. The mass of the complex scalar forming a Q--ball, the total mass of the Q--ball and its radius largely depend on the structure of the underlying theory and do not have conclusive theoretical nor observational constraints \citep{Bai_2022,SupeMBH_2014,SMBH_Gb_2016,ansariqballsSky,WickCutkosky_2017}.\\
%Dark exceptional glueball fragments can be swept away during the evolution of the galaxy, like strangelets [], assuming the current NFW profile. 

An analogous reasoning can be applied for the $GGG$ glueball made of three exceptional massive gluons, a sort of proton--like state with spin 1, if stabilizeable, introducing a vector DM--new Higgs portal \citep{Vectorportal1,Vectorportal2} and building up Proca stars \citep{Brito:2015pxa,Landea_2016,Minamitsuji_2018}. All the scenarios demonstrate that the $G(2)$ dark sector could be very versatile in reproducing dark matter phenomenology \citep{Masi2021}.
Concluding, the ways to test this enlarged $G(2)$ strong sector which includes exceptional--colored glueball DM can be summarized as follows: i) the discovery of a heavy scalar at colliders, the exotic Higgs, if it posses an albeit small interaction with the visible sector (or at least the SM Higgs), with a possible consequent decay into exceptional DM states; ii) the observation of the $G(2)-SU(3)$ phase transition produced in future colliders collisions, like the phase transition between a hadrons system and the quark--gluon plasma (QGP); iii) an astrophysics detection of an exotic object/system, such as a neutron--like star made of $G(2)$ strong matter (a "Majorana star"), glueball--stars, light PBHs or Q--balls; iv) a cosmological $G(2)-SU(3)$ first order PT \citep{Cutting_2018,PhysRevLett.115.181101,Zhou_2020,2021JHEP...05..160Z} detectable by future gravitational waves detectors in space, such as eLISA \citep{Huang_2021}, or gravitational waves from PBHs mergers \citep{Ballesteros_2022}.

%%%%%%%%%%%%%%%%%%%%%%%%%%%%%%%%%%%%%%%%%%%%%%%%%%%%%%

\section{Conclusions}
The choice of $G(2)$, as non--Abelian simply connected and center free exceptional Lie algebra of rank 2, is the minimal viable extension of the Standard Model including a reliable exotic sector, requiring no additional particle families nor extra fundamental forces. The real and complex $G(2)$ groups were deeply discussed, leading to a proper solution to enlarge the $SU(3)$ group of the SM which is not only a lattice QCD tool for computation. 
%It has been shown that $G(2)$ can be complexified into $G(2)_{\mathbb{C}}$, producing correct complex fundamental and adjoint representations and the same structural decomposition properties of real $G(2)$, due to its peculiar mathematical features. 
This represents a solid construction of a coherent particle gauge group to describe a larger strong interaction, which operated in the Early Universe before the emergence of visible matter. When the Universe cooled down, reaching a proper far beyond TeV energy scale at which $G(2)$ is broken, usual $SU(3)$ QCD appeared through a first order phase transition, while an extra Higgs mechanism produced a secluded sector of cold exceptional--colored bosons: the behavior of this PT has been analyzed from the point of view of the coupling constants running, putting in evidence the potential unification of the forces strength. It has also been shown that $G(2)$, once broken via a Higgs mechanism, is characterized by the proper color charges, anticharges and singlet states w.r.t. $SU(3)$, which are grouped in its fermionic fundamental $\{7\}$ real representation. The nature of the possible Yukawa--like interaction between this exotic Higgs and the $G(2)$ real Majorana fermions, which do not have defined baryon and lepton numbers, was discussed, pointing out the peculiarity of the $G(2)$ spectrum before and after the symmetry breaking. The SM can be naturally embedded in this framework with a minimal additional particle content, \textit{i.e.} a heavy scalar Higgs particle, responsible for a Higgs mechanism for the strong sector symmetrical w.r.t. the EW one, and a bunch of massive vector gluons whose composite states play the role of DM. Even if the extra Higgs possesses visible decay channels, it belongs to a very high energy scale which is certainly beyond current LHC searches, \textit{i.e.} a tens/hundreds of TeV scale at least, defined by its vaccum expectation value. Furthermore, the resulting exceptional complex scalar glueball DM is certainly compatible with direct, indirect, collider searches and astrophysical observations, as it is almost collisionless. Several advantageous occurrences guided the development of the theory: the mathematical correspondence with automorphism groups of Cayley--Dickson algebras \citep{Masi2021}; the great simplicity of the exceptional theory and of the new Higgs sector; the energy scale of the $G(2)-SU(3)$ PT suggested by the slope of the running constant, which could be compatible with the \textit{great desert} of particles highlighted by LHC and, at the same time, it could cure the metastability of the SM vacuum; the rich astrophysical realization which includes a suitable interplay between dark glueball--stars and BHs.
Concluding, $G(2)$ phenomenology guarantees peculiar manifestations in terms of extreme astrophysical compact objects, such as hypothetical $G(2)$ neutron--like stars, boson stars made of $G(2)$ glueballs, Q--balls or even BHs made of collapsed $G(2)$ glueball--stars, which can populate the dark halos and be disentangled from known stars in the near future.\\

\section{Acknowledgment}
I want to thank John Baez, Fiorenzo Bastianelli, Giacomo Cacciapaglia, Cristina Draper Fontanals, Brian Hall and Pietro Longhi for the very useful discussions, suggestions and explanations about the mathematical structure of $G(2)$ representations.\\
\\

\section{Appendix A}
To be as general as possible, if we consider not-necessarily-holomorphic \textit{reps}, we have to extend them holomorphically to the $G_{\mathbb{C}}$ complexification of $G$. From the Lie algebra point of view, we consider $g_{\mathbb{C}}$ as the set of formal pairs $X+\textbf{J}Y$ with $X$ and $Y$ in $g$ and $\textbf{J}$ a complex structure \citep{Hall}, distinct from the original one on $g$. $g_{\mathbb{C}}$ is isomorphic to the direct sum of two copies of $g$, in the form of $X+\textbf{i}\textbf{J}Y$ and $X-\textbf{i}\textbf{J}Y$ \citep{Hall}. If $G$ is simply connected, its "complexification" is the unique simply connected group whose Lie algebra is $g_{\mathbb{C}}$ and it is also the product of two copies of $G$; by complex linearity one obtains that $\pi(X + \textbf{J} X) = \pi(X) + \textbf{i} \pi(Y)$. 
Since complexified $G_{\mathbb{C}}$ is a product of two copies of $G$ \citep{Hall}, the irreducible \textit{reps} have the form of a tensor product of an irreducible \textit{rep} $\pi$ of the first copy of $G$ with an irreducible \textit{rep} $\pi^{'}$ of the second copy. The \textit{reps} of the first copy of $G$ inside complexified $G_{\mathbb{C}}$ are the holomorphic \textit{reps} of the original $G$, while the \textit{reps} of the second one are the anti--holomorphic \textit{reps}. From this, we conclude that the not-necessarily-holomorphic irreducible \textit{reps} of $G$ have the form of the tensor product of a holomorphic \textit{rep} of $G$ with an anti--holomorphic one: in general the tensor product of two irreducible \textit{reps} of $G$ will not be irreducible as a rep of $G$, but in the $G(2)$ special case it is \citep{Hall}.

\section{Appendix B}
The Standard Model alphas running is computed following the usual formalism \citep{manoukian2016quantum,Czakon_2005,MACHACEK198383} w.r.t. the Z boson mass scale, with $\beta=-\frac{11}{3}C_2(D) + s\frac{2}{3}n_{fl} S(R,f) + \frac{1}{3}n_H S(R,S)$ the general one--loop expression for the groups, with $s=2$ for Dirac fermions, $S(R,f)=S(R,S)=3/5\mbox{Tr}[Y^2/4]$ ($Y$ is the hypercharge operator) and $C_2(D)=0$ for $U(1)$, whereas $S(R,f)=S(R,S)=1/2$ and $C_2(D)=N$ for $SU(N)$:

\begin{align*}
\alpha^{-1}_{em}(\mu)&= \alpha^{-1}(m_Z)\frac{3}{5}cos^2\theta_W(m_Z) - \frac{\beta_{em}}{2\pi} ln\left(\frac{\mu}{m_Z}\right)\approx 59 - \frac{41}{20\pi}ln\left(\frac{\mu}{m_Z}\right)\\
\alpha^{-1}_{EW}(\mu)&= \alpha^{-1}(m_Z)sin^2\theta_W(m_Z) - \frac{\beta_{EW}}{2\pi} ln\left(\frac{\mu}{m_Z}\right)\approx 29 + \frac{19}{12\pi}ln\left(\frac{\mu}{m_Z}\right)\\
\alpha^{-1}_{S}(\mu)&= \alpha^{-1}_{S}(m_Z) - \frac{\beta_{S}}{2\pi} ln\left(\frac{\mu}{m_Z}\right)\approx 8.5 + \frac{7}{2\pi}ln\left(\frac{\mu}{m_Z}\right)
\end{align*}
where the explicit betas are
\begin{align*}
\label{alphas}
\beta_{em}&= \frac{2}{3}n_{fl} + \frac{1}{10}n_{H}\\
\beta_{EW}&= -\frac{22}{3} + \frac{2}{3}n_{fl} + \frac{1}{6}n_{H}\\
\beta_{S}&= -11 + \frac{2}{3}n_{fl}
\end{align*}
with $n_{fl}=6$ and $n_H=1$ for no BSM physics. The SM parameters fine--structure $\alpha$, Weinberg angle $\theta_W$ and strong coupling $\alpha_S$ are computed using the most up-to-date values from the \textit{Particle Data Group} (PDG).
\\
\\
All data generated or analysed during this study are included in this published article and its supplementary information files.

\bibliography{bibliography} 
\bibliographystyle{ieeetr}

\end{document}